\documentclass[conference]{IEEEtran}

\usepackage{cite}
\usepackage{listings}
\usepackage{multirow}
\usepackage{booktabs}
\usepackage{graphicx}
\graphicspath{{./images/}}
\usepackage{hyperref}
\hypersetup{colorlinks=true,linkcolor=black,urlcolor=black,citecolor=black}
\usepackage{xcolor}
\usepackage{subcaption}

\begin{document}

\title{On the Integration of Self-Sovereign Identity with TLS 1.3 Handshake to Build Trust in IoT Systems}

\author{\IEEEauthorblockN{Leonardo Perugini and Andrea Vesco}
\IEEEauthorblockA{LINKS Foundation - Cybersecurity Research Group\\
Via P.C. Boggio, 61 -- Torino 10138, Italy\\
Email: \{leonardo.perugini, andrea.vesco\}@linksfoundation.com}}

\maketitle

\begin{abstract}
The centralized PKI is not a suitable solution to provide identities in large-scale IoT systems. The main problem is the high cost of managing X.509 certificates throughout their lifecycle, from installation to regular updates and revocation. The Self-Sovereign Identity (SSI) is a decentralised option that reduces the need for human intervention, and therefore has the potential to significantly reduce the complexity and cost associated to identity management in large-scale IoT systems. However, to leverage the full potential of SSI, the authentication of IoT nodes needs to be moved from the application to the Transport Layer Security (TLS) level.  
This paper contributes to the adoption of SSI in large-scale IoT systems by addressing, for the first time, the extension of the original TLS 1.3 handshake to support two new SSI authentication modes while maintaining the interoperability with nodes implementing the original handshake protocol. The open source implementation of the new TLS 1.3 handshake protocol in OpenSSL is used to experimentally prove the feasibility of the approach.   
\end{abstract}

\IEEEpeerreviewmaketitle

\section{Introduction}
\label{sec:intro}

The Internet of Things (IoT) is the second trend driving the deployment of applications adopting Public Key Infrastructure (PKI)~\cite{ponemon}, meaning that the digital identity of these nodes is increasingly fundamental. 
It is widely believed that the adoption of a PKI with X.509 certificates \cite{rfc5280} is not the most appropriate solution for managing identities in large-scale deployments where numerous IoT nodes coexist. In this paper, we refer to well-resourced IoT and edge nodes.   
The main problem is the high cost of managing certificates throughout their lifecycle, from installation to regular updates and revocation. In addition, the centralised nature and the excess of trust into Certificate Authority (CA) make PKI vulnerable to single points of failure. The process used by a CA to issue and revoke certificates lacks transparency, the compromise of a CA serving a large system requires a complete update of all node identities before network operations can resume, and there is not a standard enrollment solution tailored to IoT nodes and protocols~\cite{8717910,10230166,9705036,10076881,
2020101658}. 

The Self-Sovereign Identity(SSI) model~\cite{SSIbook} represents an alternative solution to PKI to reduce the complexity and the related cost of X.509 certificate management in large scale IoT systems. SSI is an emerging decentralised digital identity model. It gives a node full control over the data it uses to generate and prove its identity. 
This new model relies on three fundamental elements: Distributed Ledger Technology (DLT) as the Root-of-Trust (RoT) for public identity data, Decentralized IDentifier (DID), and Verifiable Credential (VC) as the key component of the identity. Both DID~\cite{DID} and VC~\cite{VC} are in the process of being standardised by the W3C.

A node using SSI generates its identity key pair ($sk_{id}, pk_{id}$) and stores the public key $pk_{id}$ in the distributed ledger for other nodes to authenticate it. A node's DID represents the distributed ledger address where other nodes can retrieve its public key. Once these two components have been generated, a node can request a VC from one of the Issuers available in the system. The VC contains the metadata and claims about the identity of the node that holds it. The purpose of a VC is to describe the identity of the node, just like any other physical credential in our real world.
The combination of the key pair ($sk_{id}, pk_{id}$), the DID and at least one VC forms the digital identity in an SSI native IoT system. This composition of the digital identity reflects the decentralised nature of the SSI model. A node builds its own identity and, most importantly, is able to update its own identity for key rotation or any other reason in a self-sovereign manner. A node can also revoke its own identity. There is no central authority that provides all three components of identity to a node, and no central authority is able to fully revoke a node's identity. In addition, a node may enrich its identity with multiple VCs issued by different issuers. 

We believe that an SSI native IoT node reduces the need for human intervention in identity management, and therefore has the potential to significantly reduce the complexity and cost of identity management in large-scale IoT systems. 
Given this advantage, the research question underlying this paper is how to adopt SSI for authentication purposes in Transport Layer Security (TLS) v1.3~\cite{rfc8446}. Finding an adequate answer to this research question will lead to a comprehensive solution for trustworthy interactions in large-scale IoT deployments with lower associated management costs. 
Most of the discussions and proposals for authentication in the SSI framework consider an implementation at the application layer of the TCP/IP stack \cite{ssi-iot, diam-iot}. 
We believe in the implementation of the SSI authentication at the transport layer of the TCP/IP stack through and directly within the TLS v1.3 to maximize the advantages to use SSI. Integrating this new authentication mode requires extending the TLS handshake to work with VCs and DIDs in addition to X.509 certificates. 

This paper contributes to the adoption of SSI in large-scale IoT systems by addressing, for the first time, the extension of the original TLS 1.3 to support two new authentication modes with VCs and DIDs. The paper presents the novel design of the TLS 1.3 handshake protocol and proves its  ability to handle hybrid handshakes or to fall back to the original handshake. These important features enable a gradual deployment of SSI in existing IoT systems. Moreover, the new handshake protocol retains all the security features of the original one.
The paper describes the new TLS extensions and messages and their open source implementation in OpenSSL. In addition, the paper proposes a novel analytical model in order to estimate the performance of the SSI handshake with respect to the original one and to provide the research community with the evidence of the factors influencing the performance to  improve the current implementation. 
Finally, the paper discusses the results of the experimental tests to prove the feasibility of the design, the accuracy of the analytical model and to discuss future research directions.

\section{Self-Sovereign Identity (SSI)}
\label{sec:ssi}

The SSI reference framework consists of three layers. Each layer contributes to the generation of the identity and defines the basic principles for trustable interactions with the other nodes. Note that the SSI model subtends the peer-to-peer relationship between nodes in the system. 

Layer 1 is implemented by any DLT acting as a RoT for public identity data. In fact, DLTs are distributed and immutable means of storage by design~\cite{DLTs}. The Decentralized IDentifier (DID)~\cite{DID} is the new type of globally unique identifier designed to verify a node. The DID is a Uniform Resource Identifier (URI) in the form
\begin{center}
    \emph{did : method-name : method-specific-id}    
\end{center}
where \emph{method-name} is the name of the DID Method used to interact with the DLT and \emph{method-specific-id} is the pointer to the DID Document stored in the distributed ledger. Thus, DIDs associate a node with a DID Document~\cite{DID} to enable trusted interactions with it. The following is an example of a DID Document containing the DID and an Ed25519 public key used by the node for authentication purposes

\begin{footnotesize}
\begin{verbatim}
{ 
  "@context": ["https://www.w3.org/ns/did/v1"],
  "id": "did:method-name:123456789",
  "authentication": [{
    "id": "did:method-name:123456789#keys-1",
    "type": "Ed25519VerificationKey2023",
    "controller": "did:method-name:123456789",
    "publicKeyMultibase": "zH3C2AVvLMv6gmMNa
      m3uVAjZpfkcJCwDwnZn6z3wXmqPV"
  }] 
}
\end{verbatim}
\end{footnotesize}

The DID Method~\cite{DID,DID-registry} is a software implementation used by a node to interact with the DLT of choice. In accordance with W3C recommendation~\cite{DID}, a DID Method provides the functionalities to
 
\begin{itemize}
    \item \textbf{Create} a DID: generate an identity key pair ($sk_{id}, pk_{id}$) for authentication purposes, the corresponding DID Document containing the public key $pk_{id}$ and store the DID Document in the distributed ledger at the \emph{method-specific-id} pointed to by the DID, 
    \item \textbf{Resolve} a DID: retrieve the DID Document from the \emph{method-specific-id} on the ledger pointed to by the DID, 
    \item \textbf{Update} a DID: generate a new key pair ($sk_{id}', pk_{id}'$) and store a new DID Document to the same or a new \emph{method-specific-id} if the node needs to change the DID, and 
    \item \textbf{Deactivate} a DID: provide an immutable evidence in the distributed ledger that the DID has been revoked by the owner. 
\end{itemize}
The DID Method implementation is ledger-specific and makes the upper layers independent of the DLT of choice. 
\\

Layer 2 uses DIDs and DID Documents to establish a cryptographic trust between two nodes. In principle, both nodes prove the ownership of their private key $sk_{id}$ bound to the public key $pk_{id}$ in their DID Document stored in the distributed ledger.

While Layer 2 uses DID technology (i.e. the security foundation of the SSI framework) to start authentication, Layer 3 completes it and also deals with authorisation to services and/or resources using Verifiable Credentials (VCs)~\cite{VC}. A VC is a digital credential that contains additional characteristics of a node's identity beyond its identity key pair, the DID and the DID Document. 

\emph{The combination of the identity key pair, the DID, and at least one VC forms the digital identity in the SSI framework.}
\\

Layer 3 works in accordance with the classical Triangle-of-Trust. Three different roles coexist:
\begin{itemize}
  \item \textbf{Holder} is the node that owns one or more VCs and generates a Verifiable Presentation (VP) to request services and resources from a Verifier; 
  \item \textbf{Issuer} is the node that asserts claims about the identity of a node, creates a VC from those claims, and signs it before issuing the VC to the Holder;
  \item \textbf{Verifier} is the node that receives a VP from the Holder and verifies two signatures, one made by the Issuer on the VC and one computed by the Holder on the VP, before granting or denying the access to a service or a resource based on the claims in the VC.
\end{itemize}

The Issuer signs the VC to make it a verifiable digital credential. The Holder requests access to services and resources from a Verifier by presenting a VP. The Holder builds the VP as an envelope of the VC and signs it with its identity private key $sk_{id}$. The Issuers are also responsible for revoking VCs for cryptographic integrity and/or for status change purposes~\cite{VC}.
\\

The VC contains the metadata to describe properties of the credential (e.g. context, id, type, issuer, issuance and expiration dates) and most importantly, the DID and the claims about the identity of the node in the \verb+credentialSubject+ field.

The following is an example of VC of type \verb+IoTCredential+ for an IoT node issued and signed by the Issuer identified by its DID \verb+did:method-name:abcdefghi+; refer to~\cite{VC} for a detailed explanation of each field.

\begin{footnotesize}
\begin{verbatim}
{
  "@context": 
    ["https://www.w3.org/2018/credentials/v1"],
  "id": "https://address/credentials/1",
  "type": ["VerifiableCredential", "IoTCredential"],
  "issuer": "did:method-name:abcdefghi",
  "issuanceDate": "2023-09-19T15:34:40Z",
  "expirationDate": "2025-01-01T12:00:00Z",
  "credentialSubject": {
    "id": "did:example:123456789",
    .. properties to describe the identity ..
  },
  "proof": {
    "type":	"Ed25519VerificationKey2023",
    "created":	" 2023-09-19T15:34:40Z",
    "proofPurpose":	"assertionMethod",
    "verificationMethod": "did:method-name:abcdef
      ghi#key-1",
    "proofValue": .. the signature ..
  }
}
\end{verbatim}
\end{footnotesize}

\section{Transport Layer Security (TLS)}
\label{sec:tls}

\subsection{Original handshake}
\label{sec:original-handshake}

\begin{figure*}[t]
    \begin{subfigure}{0.5\textwidth}   
        \centerline{\includegraphics[width=\columnwidth]{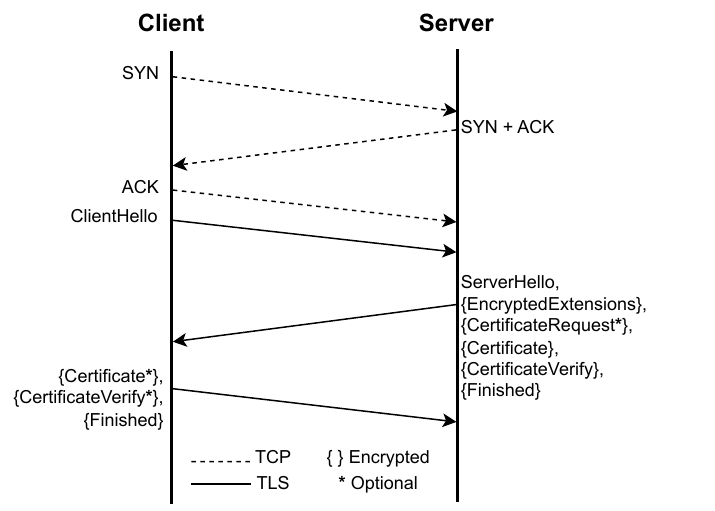}}
        \caption{}
        \label{fig:original-tls-hs}
    \end{subfigure}
    \begin{subfigure}{0.5\textwidth}
        \centerline{\includegraphics[width=\columnwidth]{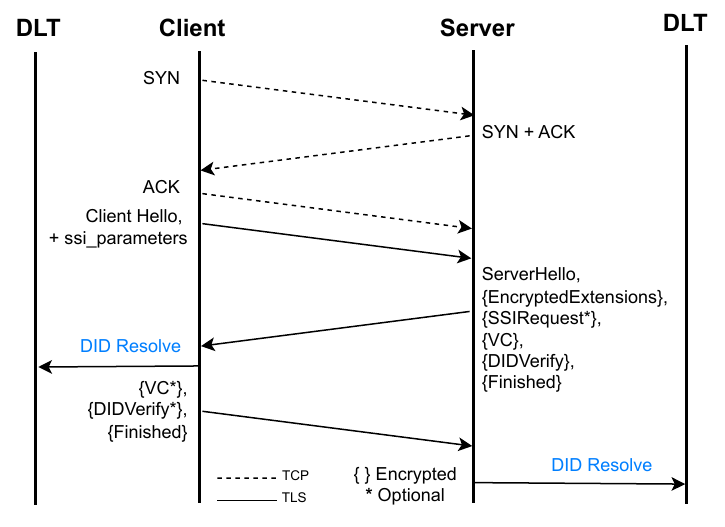}}
        \caption{}
        \label{fig:ssi-tls-hs}
    \end{subfigure}
\caption{Message flow in (a) the original TLS 1.3 handshake protocol and (b) the SSI TLS 1.3 handshake protocol.}
\label{fig:original-and-ssi-hs}
\end{figure*}

The TLS provides a secure channel between a client and server over the Internet. The secure channel provides server (and optionally client) authentication, confidentiality and integrity of messages in transit. 
The TLS 1.3 handshake protocol establishes the secure channel by exchanging the messages shown in Figure~\ref{fig:original-tls-hs}. 

Upon the handshake, client and server negotiate cryptographic parameters through the exchange of \texttt{ClientHello} and \texttt{ServerHello} messages. Those parameters are the symmetric cipher and hash algorithm used to ensure confidentiality and integrity respectively. 
The client and the server generate a secret by using ephemeral Diffie–Hellman(DH) key exchange and use the HMAC-based Extract-and-Expand Key Derivation Function (HKDF) algorithm with the negotiated hash algorithm to derive the session keys from the secret. 
Then, the server authenticates with the client by sending the \texttt{Certificate} message containing the certificate chain (Root CA certificate excluded), and the \texttt{CertificateVerify} message which is the signature over all the exchanged messages computed with its private key. 
The client verifies the validity of the certificate chain and the signature in the \texttt{CertificateVerify} message to check the identity of the server. 
The server can request client authentication by sending the \texttt{CertificateRequest} message. The client authenticates with the server in the same way. 

\begin{figure*}[t]
    \begin{subfigure}{0.5\textwidth} 
    \centerline{\includegraphics[width=\columnwidth]{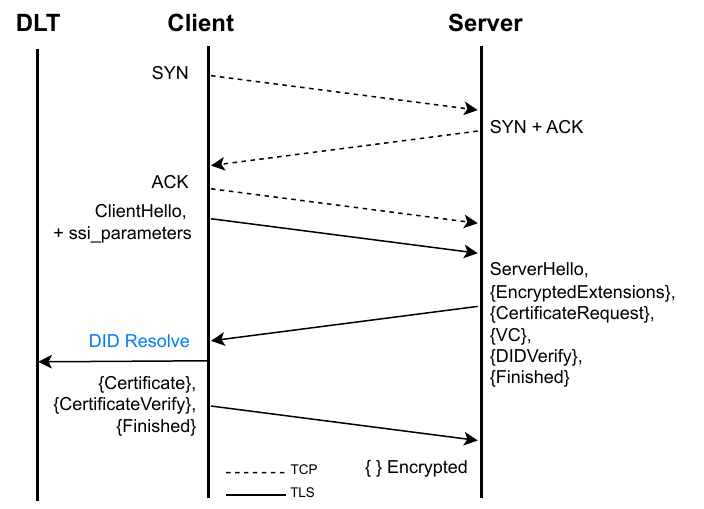}}
        \caption{}
        \label{fig:hybrid-hs-1}
    \end{subfigure}
\begin{subfigure}{0.5\textwidth}            
    \centerline{\includegraphics[width=\columnwidth]{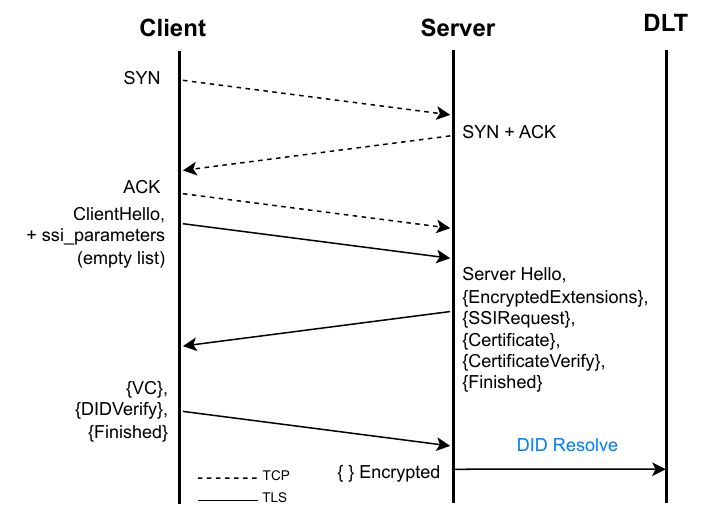}}
    \caption{}
    \label{fig:hybrid-hs-2}
\end{subfigure}
\caption{Hybrid TLS 1.3 handshakes with (a) original client and SSI server and (b) SSI client and original server.}
\label{fig:hybrid-hs-1-2}
\end{figure*}

\subsection{SSI handshake} 
\label{sec:ssi-handshake}

Our goal is threefold, (\emph{i}) design a new authentication mode that allows a server (and optionally a client) to authenticate using a VC, (\emph{ii}) preserve the interoperability with public key certificates, i.e. support  hybrid handshake with certificate and VC and a fallback to the original handshake, and (\emph{iii}) retain all the security features of TLS 1.3. Figure~\ref{fig:ssi-tls-hs} shows the flow of messages with the new extensions of the new handshake protocol.
The protocol uses a brand new TLS extension called \texttt{ssi\_paramaters} that allows client and server to negotiate the VC authentication mode and a common set of DID Methods for resolving DIDs (see DID Resolve function in Section~\ref{sec:ssi}).
The client sends this extension with the \texttt{ClientHello} to start an SSI handshake. A server can request client authentication through the \texttt{SSIRequest} message with the \texttt{ssi\_parameters} extension as well. 

As discussed in Section~\ref{sec:ssi}, a Holder wraps the VC into a VP and signs it before presenting it to a Verifier for authentication. The SSI handshake captures this design principle by means of two new messages: the \texttt{VC} message that carries the VC and the \texttt{DIDVerify} message that is a signature performed over all messages and the VC with the server private key ($sk_{id}^s$). The randomness in the signature consist of all handshake messages before the \texttt{DIDVerify} message. This design choice reduces the total number of signatures and signature verifications during the handshake while maintaining compliance with the VP concept.

Upon receiving \texttt{DIDVerify} message, the client check that the VC follows the scheme specified in the \texttt{@context} field, checks the validity of the VC metadata, verify the signature of the Issuer on the VC, and then extract the server DID from the \verb+credentialSubject+ field of the VC and resolve the server DID to retrieve the server public key $pk_{id}^s$ from the distributed ledger. Finally, the client verifies the signature in the \texttt{DIDVerify} message. Notice, both client and server maintain a list of the public keys of Issuers they trust. The client authenticates with the server in the same way.

\subsection{Hybrid handshake}
\label{sec:hybrid-handshake}

The new extensions and messages introduced in Section~\ref{sec:ssi-handshake} made possible to use VC as a new authentication mode without modifying the original handshake that use public key certificates as authentication mode. This design choice leaves room for the gradual introduction of SSI in existing systems.

In fact, it is still possible to authenticate both client and server with a hybrid handshake if either the client or the server wants to authenticate with an X.509 certificate, while the other wants to use VC. This option originates two different handshake flavours shown in Figure~\ref{fig:hybrid-hs-1-2} %

In both cases, the client sends the \texttt{ssi\_parameters} extension with the \texttt{ClientHello} to start an SSI handshake. 
In the first case, the client sends the list of DID Methods it supports to request the server to authenticate with a VC, whereas the server sends back a \texttt{CertificateRequest} message to request the client to authenticate with an X.509 certificate. 
In the second case, the client sends an empty list of DID Methods to inform the server it possesses a VC, but wants the server to authenticate with an X.509 certificate. Accordingly, the server sends in sequence \texttt{SSIRequest} message with the whole set of DID Methods it supports since the client did not send any, \texttt{Certificate} and \texttt{CertificateVerify} messages.

\subsection{Fallback mechanism}
\label{sec:fallback}

A fallback mechanism to the original handshake is needed for interoperability purposes. A server could always decide to fall back to the original handshake when the client requests an SSI handshake. 
In the case a client sends the \texttt{ssi\_parameters} extension to a server that does not support SSI authentication mode, the latter ignores the extension and proceeds by sending in sequence \texttt{Certificate} and \texttt{CertificateVerify} messages back to the client. In case the server requests client authentication, the client must proceed with the original handshake.
In the case a server supports VC authentication mode, but does not own a DID in one of the DLT specified by the client in the list of DID Methods in the \texttt{ssi\_parameters} extension, the server again falls back to the original handshake.

\section{Detailed Design of the SSI Handshake}
\label{sec:design}

The extension and message formats described in the following subsections retain the same syntax as in RFC8446~\cite{rfc8446}.

\subsection{Extensions}

\subsubsection{ssi\_parameters} extension is allowed only in \texttt{ClientHello} and \texttt{SSIRequest} messages, and it is meant to trigger the SSI handshake. The structure of the extension is as follows:

\begin{verbatim}
enum {
    0, DID(1), VC(2)
} AuthenticationMode
enum {
    iota(0), btcr(1), …, (255)
} DIDMethod
Struct {
      AuthenticationMode authn;
      DIDMethod did_methods<1..2^8-1>;
} SSIParameters;
\end{verbatim}

The \texttt{AuthenticationMode} field can be either DID or VC. Further discussion on DID authentication mode is presented in Section~\ref{sec:did-handshake}. 
The second field is a list of \texttt{DIDMethod}, namely the list of DLT the endpoint is able to interact with. Note that a node creates a DID in at least one of the DLT it supports.
Each DID Method is mapped to a single byte integer. 
Any endpoint processing this extension must check that its DID belong to one of the DID Methods in the list, otherwise the other endpoint will never be able to resolve it to retrieve the corresponding public key $pk_{id}$ from the distributed ledger. 
In order to support the hybrid handshake discussed in Section~\ref{sec:hybrid-handshake} and represented in Figure~\ref{fig:hybrid-hs-2} the integer byte in the \texttt{AuthenticationMode} field is set to 0 and the list in \texttt{DIDMethod} is left empty.
\\

\subsubsection{signature algorithms} defined in RFC8446~\cite{rfc8446} are different from the ones proposed by W3C~\cite{DID-registry} as expected. 
In order to align the different cipher suites for the purpose of this work, we have kept the keys and signature algorithms from RFC8446~\cite{rfc8446}, and we have defined three new suites maintaining the same nomenclature proposed by the W3C recommendation as in Table \ref{tab:cyphersuites}.
It is advisable to harmonise the IETF and W3C specifications in the near future to support the use of SSI in TLS 1.3 handshake.
\\

\begin{table}[h!]
    \centering
    \begin{tabular}{cc}
    \toprule
    RFC 8446 & W3C \\
    \midrule
    ecdsa\_secp256r1\_sha256 & EcdsaSecp256r1Signature2023 \\ 
    rsa\_pss\_rsae\_sha256 & RsaSignature2023 \\ 
    ed25519 & Ed25519Signature2023 \\ 
    \bottomrule
    \end{tabular}
    \caption{Binding between cipher suites.}
    \label{tab:cyphersuites}
\end{table}

\subsection{Messages}

\subsubsection{SSIRequest} message is sent only by the server to request client authentication during SSI handshake. It contains \texttt{ssi\_parameters} and \texttt{signature\_algorithms} extensions, which must always be present. 
The first one must select the same authentication mode as the client via \texttt{ssi\_parameters} and a set of DID Methods client and server have in common.
The client must abort the handshake during the processing of \texttt{SSIRequest} message if one of the two previous conditions is not satisfied or the client does not have a DID in the distributed ledgers the server can interact with. 
The \texttt{signature\_algorithms} extension contains the list of signature algorithms the server can use to verify the signature in the \texttt{DIDVerify} message sent by the client. The structure of this message is as follows:

\begin{verbatim}
struct {
      Extension extensions<2..2^16-1>;
} SSIRequest;
\end{verbatim}

\subsubsection{VC} message carries the content of the VC. The server send this message back to the client when the latter propose the VC authentication mode. The client send this message upon receiving \texttt{SSIRequest} message with the \texttt{ssi\_parameters}
extension proposing VC authentication mode. An endpoint receiving the \texttt{VC} message process the VC as discussed in Section~\ref{sec:ssi-handshake}. The structure of this message is as follows:

\begin{verbatim}
struct {
    opaque vc<0..2^16-1>
} VC;
\end{verbatim}

\subsubsection{DIDVerify} message allows an endpoint to prove the possession of the private key $sk_{id}$ and it must be sent right after a \texttt{VC} message. It carries the signature of all previous handshake messages computed with $sk_{id}$. For the sake of clarity, $sk_{id}$ is the private key associated to the public key $pk_{id}$ in the endpoint's DID document (e.g. \verb+#keys-1+ in the example of DID Document in Section~\ref{sec:ssi}). The server must always send this message, the client sends it only when is requested to authenticate by the server. 

The structure of this message remains the same as the \texttt{CertificateVerify} message~\cite{rfc8446}, thus it contains the algorithm used for the signature plus the signature itself. The signature is computed over the concatenation of octet 32 (0x20) repeated 64 times, the context string which is “\emph{TLS 1.3, server DIDVerify}” on server side and “\emph{TLS 1.3, client DIDVerify}” on client side, a single 0 byte that acts as a separator and the hash of all previous handshake messages at the end. The structure of this message is as follows:

\begin{verbatim}
struct {
    SignatureScheme algorithm;
    opaque signature<0..2^16-1>	
} DIDVerify;
\end{verbatim}
\vspace{+3mm}

\section{DID Authentication Mode} 
\label{sec:did-handshake}

It is also worth to discuss another possible authentication mode leveraging the DIDs. In practice, the server (and optionally the client) present its DID instead of the VC. Since DIDs are self-issued and DLT are permissionless in principle, any node can store his DID Document on the distributed ledger. Therefore, the DID authentication mode requires each endpoint to maintain a list of trusted DID for authentication purposes (i.e. maintaining the list of DIDs in a trusted group). An efficient Merkle tree based solution to this membership problem, suitable for IoT systems, has been proposed in~\cite{pino-2023}.

The handshake with DID authentication mode can be triggered by selecting \texttt{DID} authentication mode in the \texttt{ssi\_parameters} extension. The flow of messages remains the same depicted in Figure~\ref{fig:ssi-tls-hs}, besides the \texttt{VC} message that gets replaced by the \texttt{DID} message. The client processing this message must first check that the DID Method to resolve the DID is present in its list of DID Methods previously sent in the \texttt{ssi\_parameters} extension, and then resolve the DID into the corresponding DID Document to extract the public key that will later be needed to verify the \texttt{DIDVerify} message. In this case, the client has to verify the trust in the DID (i.e. the DID belong to a trusted list of DID) and only the signature in the \texttt{DIDVerify} message. This option reduces by one the number of signature verification with respect to the VC authentication mode.
The \texttt{DID} message contains a byte to specify the DID Method followed by the actual DID as shown in the following:

\begin{verbatim}
struct {
    DIDMethod did_method;
    opaque did<0..2^16-1>;
} DID
\end{verbatim}

The DID authentication mode can also be employed when client authentication is requested by the server and in hybrid handshake cases.

\section{Security Analysis}
\label{sec:security}

The proposed SSI handshake works with TLS 1.3 therefore it is important to retain the same security properties of the original handshake. Here we consider a Dolev-Yao attacker \cite{dolev1983security} that has full control of the network and can intercept, send, replay and delete any message. Moreover, we assume that an attacker can encrypt and decrypt messages if it knows the appropriate keys. 

In the SSI handshake, the way session keys are established remains the same, thus supplying perfect forward secrecy and the same level of confidentiality and integrity as in the original handshake. All the messages after \texttt{ServerHello} are encrypted with the handshake session keys. The authentication of the endpoints is performed by the combination of \texttt{VC} and \texttt{DIDVerify} messages. 
This follows the same asymmetric challenge-response mechanism proposed in the original handshake with the \texttt{Certificate} and \texttt{CertificateVerify} messages. 
The challenge corresponds to the transcript-hash as in the original handshake. Thus, an attacker is not able to impersonate an authenticating endpoint unless it discovers its long term secret key. 
The \texttt{ssi\_parameters} in \texttt{ClientHello} is sent in clear, and this is acceptable since it does not contain any confidential information. If the client is requested to authenticate, the \texttt{ssi\_parameters} extension benefits from the integrity and authentication property provided respectively by the \texttt{Finished} and \texttt{DIDVerify} messages, otherwise only its integrity is guaranteed. 
Since the server always authenticate with the client, the \texttt{ssi\_parameters} extension is sent in the \texttt{SSIRequest} message and benefits from both properties.

Some considerations about the DID resolution process are worth discussing. Assuming that a DID resolution is performed in clear, the same attacker could impersonate the DLT node, forge a DID document containing the authenticating endpoint's DID, associate it with a key pair that he owns, and then return it to the DID resolver. Thus, the attacker is able to compute a valid \texttt{DIDVerify} message by possessing the long term private key. In practice, the man-in-the-middle attacker breaks in transit the immutability feature of the DLT (i.e. the RoT for identity public keys).
A reasonable solution to this attack could be to create a TLS channel towards the DLT node and authenticate only the latter to rely on the received data. The DLT node must be authenticated through an X.509 certificate. The number of DLT nodes within an IoT large scale systems is expected to be very low (i.e. one or a couple of nodes) with respect to the total number of IoT and edge nodes, so adopting X.509 certificates to authenticate those DLT nodes does not reduce the overall benefit in terms of lower complexity and cost associated to certificate management proper of SSI solution.  
In order to reduce the overhead of establishing a TLS channel with the DLT node for DID resolution, there are two possible approaches (\emph{i}) leverage zero round trip time resumption (0-RTT) or (\emph{ii}) changing the logic of DLT nodes and adopt a data protection solution (e.g. with HMAC to authenticate the data from DLT node).

\section{Implementation in OpenSSL}
\label{sec:implementation}

To experimentally evaluate the SSI handshakes, we have implemented the new extensions and messages in OpenSSL, a globally adopted open source cryptographic library. It is mainly written in C language and consists of two sub-libraries: \textit{ssl} that implements the SSL and TLS protocol and \textit{crypto} that supplies a wide variety of cryptographic operations. 
In practice, we have designed and developed from scratch a loadable module in the form of an OpenSSL provider. This provider, called \textit{ssi}, implements all functions to deal with DID, DID Documents and VCs in accordance with their description in Section~\ref{sec:ssi}. In detail, the \textit{ssi} provider supplies the implementation of two brand-new operations: \texttt{OP\_DID} and \texttt{OP\_VC}. 

The \texttt{OP\_DID} defines the functions of a DID Method to create, resolve, update and deactivate DIDs from within OpenSSL. In this work, we adopted the specific implementations of the OTT~\cite{claudio2023} DID Method that interact with IOTA Tangle distributed ledger~\cite{Tangle}. The \texttt{OP\_DID} defines all four functionalities. In particular, the Resolve function is used in the new handshake, while the other three functions are available through the OpenSSL application layer to allow any node to create, update, and deactivate its own decentralised identity. Moreover, it is worth noting that any other DID Method can be added to the \texttt{OP\_DID} as per philosophy of provider and operation mechanisms.

The \texttt{OP\_VC} defines four functions as well, to create, verify, serialize and deserialize a VC. These functionalities are called during different portion of the overall SSI handshake. The \emph{Verify} function is called to process the \texttt{VC} message. The \emph{Serialize} and \emph{Deserialize} functions are called before sending and after receiving the \texttt{VC} message. The \emph{Creation} functionality is made available through OpenSSL application layer for testing purposes; in a real deployment scenario the node receives the VC from an Issuer.

Providers and OpenSSL communicate without knowing each other's internal structure thanks to the \textit{core} component of the \textit{crypto} library that supplies public data structures to exchange information. Therefore, the applications that want to employ OpenSSL cryptographic operations can interact with the \textit{crypto} library through the \texttt{EVP} API without calling the implementations supplied by the provider directly. \texttt{EVP} methods internally invoke the provider functions. So we added two elements to the \texttt{EVP} interface that we called \texttt{EVP\_VC} and \texttt{EVP\_DID}.

Finally, in the \textit{ssl} library we have updated the TLS 1.3 handshake state machine to use \texttt{SSIRequest}, \texttt{DID}, \texttt{VC} and \texttt{DIDVerify} messages and the \texttt{ssi\_parameters} extension. Internally, the new messages invoke \texttt{EVP\_DID} and \texttt{EVP\_VC} APIs to process DIDs, DID Documents, and VCs. 

The implementation of the \emph{ssi} provider~\cite{ssi-provider} and of the SSI handshake protocols~\cite{ssi-openssl} in OpenSSL are available in open-source.

\section{Performance Analysis}

\subsection{Analytical model}
\label{sec:model}

We have devised a theoretical model to estimate the latency of an SSI and hybrid handshake starting from the original one. In the following definitions and equations a single \('\) denotes a unilateral authenticated handshake, whereas two \(''\) denotes a mutual authenticated handshake. 
Let \(H_O^{'}\) be the average latency of an original handshake, and let \(H_V^{'}\) be the average latency of an SSI handshake with VC authentication mode, and let \(T_V\), \(T_C\) and \(T_D\) be the average time to verify a VC, verify a certificate chain and resolve a DID respectively. Starting from these definitions, the average latency of an SSI handshake with VC authentication mode can be estimated as it follows: 

\begin{equation} 
\label{eq:hv-1}
    H_V^{'} = H_O^{'} - T_C + (T_V + T_D)
\end{equation}

Equation~(\ref{eq:hv-1}) states that the latency of the two handshakes differ from each other for the processing time of the identity messages such as \texttt{Certificate} and \texttt{VC} plus the average time to resolve the server's DID and retrieve the identity public key from the distributed ledger.

Similarly, the average latency of a handshake with DID authentication mode \(H_D^{'}\) can be estimated as it follows:    

\begin{equation} 
\label{eq:hd-1}
    H_D^{'} = H_O^{'} - T_C + T_D
\end{equation}

Now, let \(\Delta_V\) and \(\Delta_D\) store the values of the distinguishing factors in the equations~(\ref{eq:hv-1}) and (\ref{eq:hd-1}), namely  

\begin{equation} 
\label{eq:delta_v}
    \Delta_V = T_V + T_D - T_C
\end{equation}

\begin{equation} 
\label{eq:delta_d}
    \Delta_D = T_D - T_C
\end{equation}

Thus, the latency of mutually authenticated SSI handshakes \(H_V^{''}\) and \(H_D^{''}\) can be estimated as it follows:

\begin{equation} 
\label{eq:hv-2}
    H_V^{''} = H_O^{''} + 2\Delta_V
\end{equation}

\begin{equation} 
\label{eq:hd-2}
    H_D^{''} = H_O^{''} + 2\Delta_D
\end{equation}

Finally, let \(H_{OV}^{''}\) and \(H_{OD}^{''}\) represent the average latency of hybrid handshakes. The first one involves the adoption of X.509 certificates at client side and a VC at server side, whereas the second involves the adoption of X.509 certificates at client side and of a DID at server side. They can be estimated as it follows:

\begin{equation} 
\label{eq:hybrid-v}
    H_{OV}^{''} = H_O^{''} + \frac{H_{V}^{''} - H_O^{''}}{2} = H_O^{''} + \Delta_V
\end{equation}

\begin{equation} 
\label{eq:hybrid-d}
    H_{OD}^{''} = H_O^{''} + \frac{H_{D}^{''} - H_O^{''}}{2} = H_O^{''} + \Delta_D 
\end{equation}
moreover, \(H_{VO}^{''} = H_{OV}^{''}\) and \(H_{DO}^{''} = H_{OD}^{''}\) are satisfied.

\subsection{Experimental Setup}
\label{sec:setup}

\begin{figure}[h!]
    \centering
    \includegraphics[width=\columnwidth]{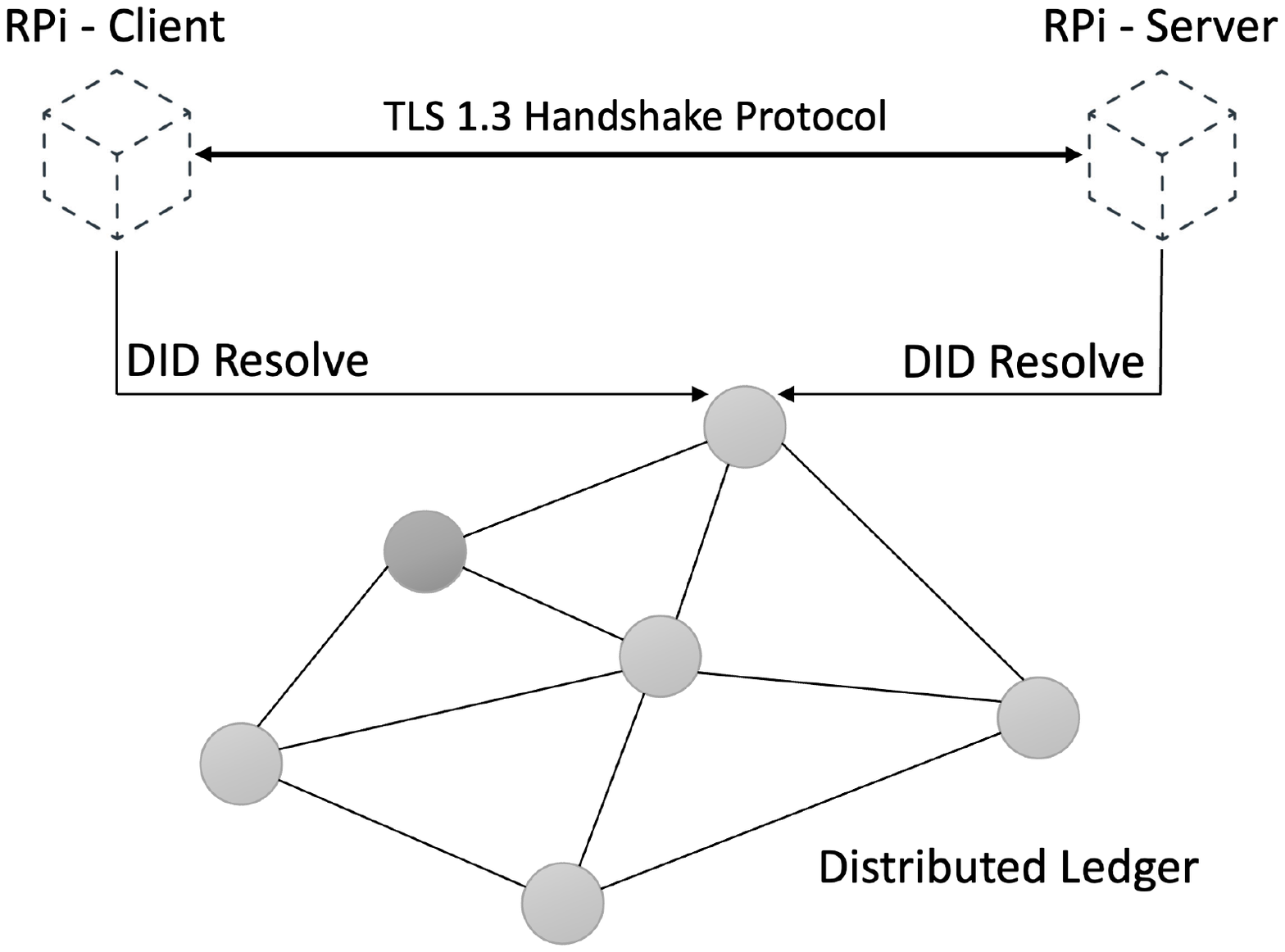}
    \caption{Experimental setup}
    \label{fig:setup}
\end{figure}

To assess the performance of the new SSI and hybrid handshakes we have installed the modified version of OpenSSL on two Raspberry Pi's 4 Model B equipped with a quad-core Cortex-A72 (ARM v8) SoC clocked at 1.8GHz, 4 GB of SDRAM, a Gigabit Ethernet interface, and 32-bit OS. The RPIs are connected in a client-server configuration and both have access to a DLT node to resolve the DIDs as depicted in Figure~\ref{fig:setup}.

In this work, the RPIs leverage the immutability feature of the IOTA Tangle and interact with it through an IOTA node installed in our lab. The RPIs authenticate the IOTA node with an X.509 certificate signed with an ECDSA 256-bit key during an original handshake, as discussed in Section~\ref{sec:security}.

In all our tests client and server adopt x25519 elliptic curve for EcDHE key exchange and \texttt{TLS\_AES\_256\_GCM\_SHA384} cipher suite. 
We have tested the performance of unilaterally authenticated handshake and mutually authenticated handshakes also in hybrid scenarios under different authentication modes (i.e. three-link X.509 certificate chain, VC and DID). Moreover, we have tested the performance for different signature algorithms, see in Table~\ref{tab:cyphersuites}.
We have run 1000 handshakes for each configuration using \texttt{s\_client} and \texttt{s\_server} applications provided by OpenSSL to collect statistically relevant results in terms of handshake size and latency. 
Client and server select randomly an X.509 certificate chain, VC or DID from a predefined large set at each run.

\subsection{Experimental results}
\label{results}

\subsubsection{handshake size} Table \ref{tab:hs-size} shows the number of bytes sent by the server to the client in a unilaterally authenticated handshake under different configurations of authentication modes and signature algorithms.
With the exception of DIDs, which have a fixed length (e.g. 72 bytes in the case of the IOTA distributed ledger), the total length of X.509 certificates and VCs can vary, therefore the total bytes column should be considered as indicative. Instead, the values about public key objects are fixed and lead to some interesting considerations. When using RSA, the server sends a larger amount of data over the network. Conversely, \texttt{ECDSA} and \texttt{EdDSA} cause the server to transmit considerably fewer data. The X.509 authentication mode requires the most bytes of public key objects to be sent. DID authentication mode is highly beneficial because the server only sends half the amount of bytes compared to VC authentication mode. In a unilaterally authenticated handshake, the server does not interact with the DLT node. Only the client resolves the server DID. 
Conversely, in a mutually authenticated SSI handshake, the server sends an additional 377 bytes to create a secure connection with the DLT node. Despite these extra bytes, DID and VC authentication modes still remain more advantageous than X.509 in terms of bytes sent over the network. IoT nodes may find these alternatives, combined with elliptic curve digital signatures, attractive.  

\begin{table}[t]
    \centering
    \begin{tabular}{llc|cccc}
        \toprule
        \multirow{2}{*}{} & \multirow{2}{*}{} & \textbf{Total Bytes} & \multicolumn{3}{c}{\textbf{Public Key Objects}} \\
        {} & {} & {} & pk & signature & tot \\
        \midrule
        \multirow{3}{*}{\rotatebox[origin=c]{90}{\textbf{X.509}}} & RSA-2048 & 2063 & 2*272 & 3*256 & 1312 \\
        {} & ECDSA & 1082 & 2*33 & 3*70 & 276 \\
        {} & EdDSA & 950 & 2*32 & 3*64 & 256 \\
        \midrule
        \multirow{3}{*}{\rotatebox[origin=c]{90}{\textbf{VC}}} & RSA-2048 & 1516 & / & 2*256 & 512 \\
        {} & ECDSA & 1094 & / & 2*70 & 140 \\
        {} & EdDSA & 1072 & / & 2*64 & 128 \\
        \midrule
        \multirow{3}{*}{\rotatebox[origin=c]{90}{\textbf{DID}}} & RSA-2048 & 623 & / & 1*256 & 256 \\
        {} & ECDSA & 437 & / & 1*70 & 70 \\
        {} & EdDSA & 431 & / & 1*64 & 64 \\
        \bottomrule
    \end{tabular}
    \caption{Unilaterally authenticated TLS 1.3 handshake size (unit: bytes)}
    \label{tab:hs-size}
\end{table}

\begin{table}[t]
    \centering
    \begin{tabular}{cc|ccc}
        \toprule
        {} & \textbf{Signature Verify} & \textbf{Certificate} & \textbf{VC} & \textbf{DID} \\
        \midrule
        RSA & 1 & 8 & 45 & 47 \\
        ECDSA & 1 & 9 & 48 & 47 \\
        EdDSA & 11 & 27 & 57 & 45 \\
        \bottomrule
    \end{tabular}
    \caption{{Average time to verify a single signature and to process identity messages. (unit: ms)}}
    \label{tab:identity-verification}
\end{table}

\begin{table}[t]
    \centering
    \vspace{+19mm}
    \begin{tabular}{lccc|ccc}
        \toprule
        \multirow{2}{*}{} & \multicolumn{3}{c}{\textbf{Unilateral Authentication}} & \multicolumn{3}{|c}{\textbf{Mutual Authentication}} \\
        {} & X.509 & VC & DID & X.509 & VC & DID \\
        \midrule
        RSA-2048 & 78 & 110 & 109 & 116 & 190 & 187 \\
        ECDSA & 43 & 78 & 77 & 50 & 124 & 122 \\
        EdDSA & 71 & 93 & 83 & 104 & 161 & 143 \\
        \bottomrule
    \end{tabular}
    \caption{Experimental average measures of the handshake latency at server side. (unit: ms)}
    \label{tab:symmetric-latency}
\end{table}

\begin{table}[t]
    \centering
    \begin{tabular}{l|cc|cc}
        \toprule
        \textbf{Server} & X.509 & X.509 & VC & DID \\
        \textbf{Client} & VC & DID & X.509 & X.509 \\
        \midrule
        RSA & 152 & 150 & 151 & 149 \\
        ECDSA & 87 & 86 & 83 & 81 \\
        EdDSA & 131 & 122 & 129 & 118 \\
        \bottomrule
    \end{tabular}
    \caption{Experimental average measures of the hybrid handshake latency at server side. (unit: ms)}
    \label{tab:asymmetric-latency}
\end{table}

\subsubsection{verification of identity messages} 
\label{sec:identity-verification}

Table~\ref{tab:identity-verification} shows the average time to verify a single signature in the first column and the average time needed to process and to verify \texttt{Certificate}, \texttt{VC} and \texttt{DID} messages in the following columns. 
Verifying the chain of certificates is quicker than verifying VCs and DIDs. In other words, it takes less time to trust and retrieve the public key of the other endpoint using certificates than it does with SSI. The reason being the requirement of creating a secure channel with the DLT node to resolve the DID of the other endpoint, which takes about 33ms. Since there are options to reduce this delay as discussed in Section~\ref{sec:security} (e.g. 0-RTT handshake), the values for VCs and DIDs in Table~\ref{tab:identity-verification} represent the worst case. 
Comparing the verification times of VC and DID, it can be seen that with RSA, the times are about the same because, although the verification of a VC requires the verification of another signature (i.e. the Issuer's signature), this verification time is negligible compared to the total time. This is not true when the signature verification time increases. In case of EdDSA the use of DIDs gives advantage over the use of VCs. This is an interesting result in view of the adoption of Post-Quantum Cryptography (PQC) in TLS 1.3. The signature verification can take longer with PQC. 

\begin{figure*}[t]
\begin{subfigure}{0.5\textwidth}
    \includegraphics[width=\columnwidth]{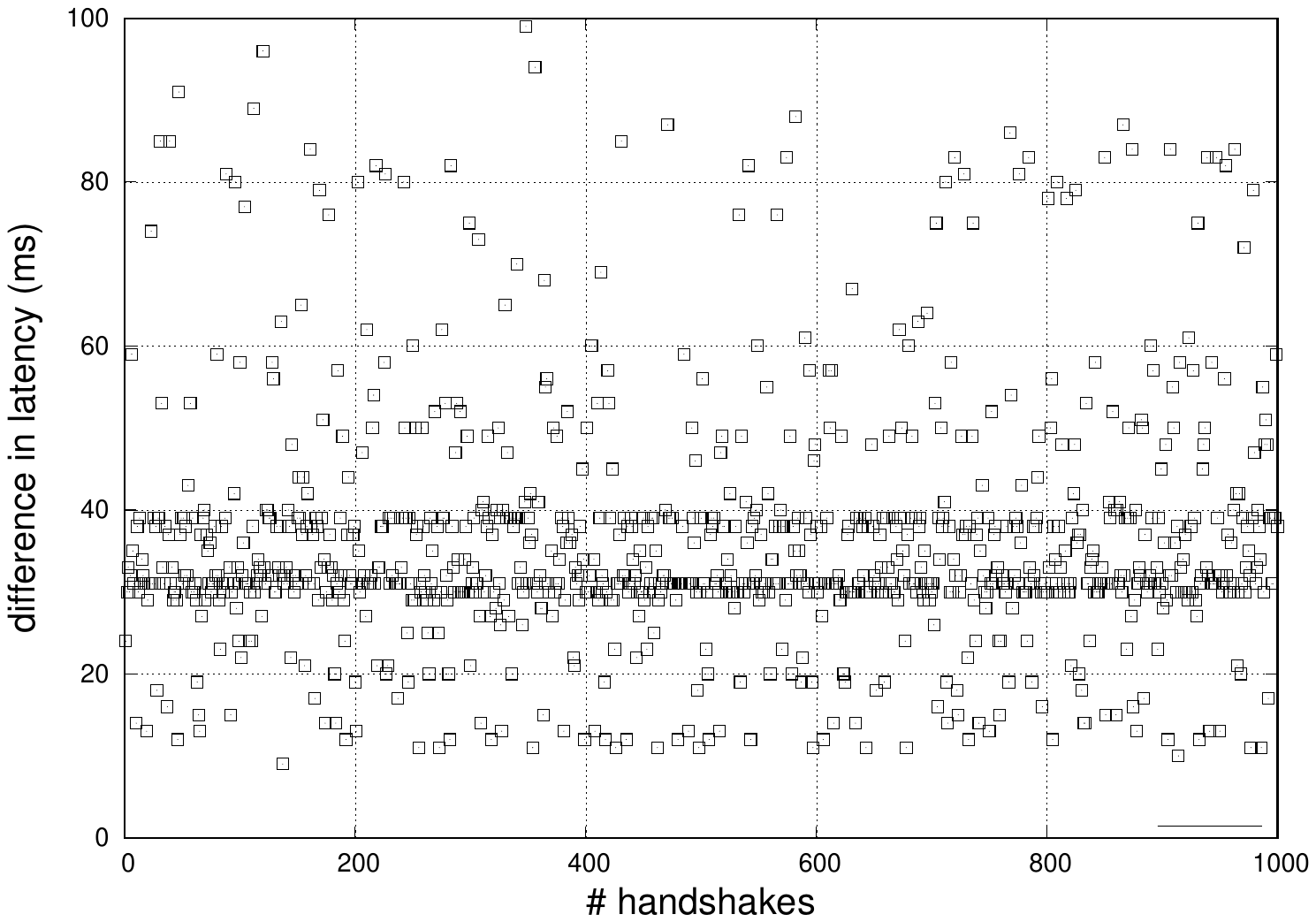}
    \caption{}
    \label{fig:one-way-measures}
\end{subfigure}
\begin{subfigure}{0.5\textwidth}
    \includegraphics[width=\columnwidth]{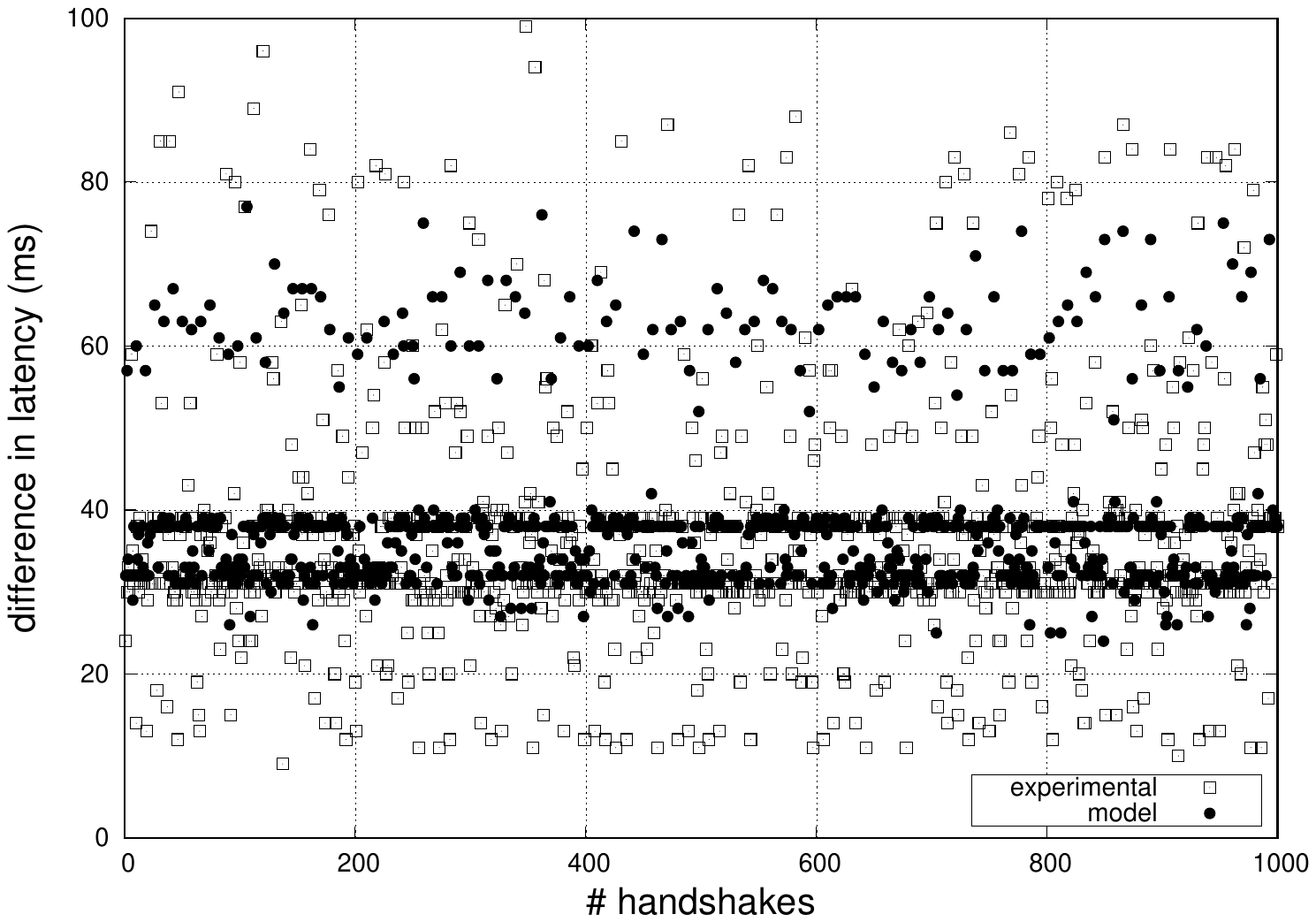}
    \caption{}
    \label{fig:one-way-model}
\end{subfigure}
\caption{(a) Experimental difference between the unilaterally authenticated handshake latency with VC and X.509 authentication modes and ECDSA signature algorithm, (b) Overlay of experimental values and model estimates.}
\label{fig:one-way-measures-model}
\end{figure*}

\begin{figure*}[t]
    \begin{subfigure}{0.5\textwidth}
        \includegraphics[width=\columnwidth]{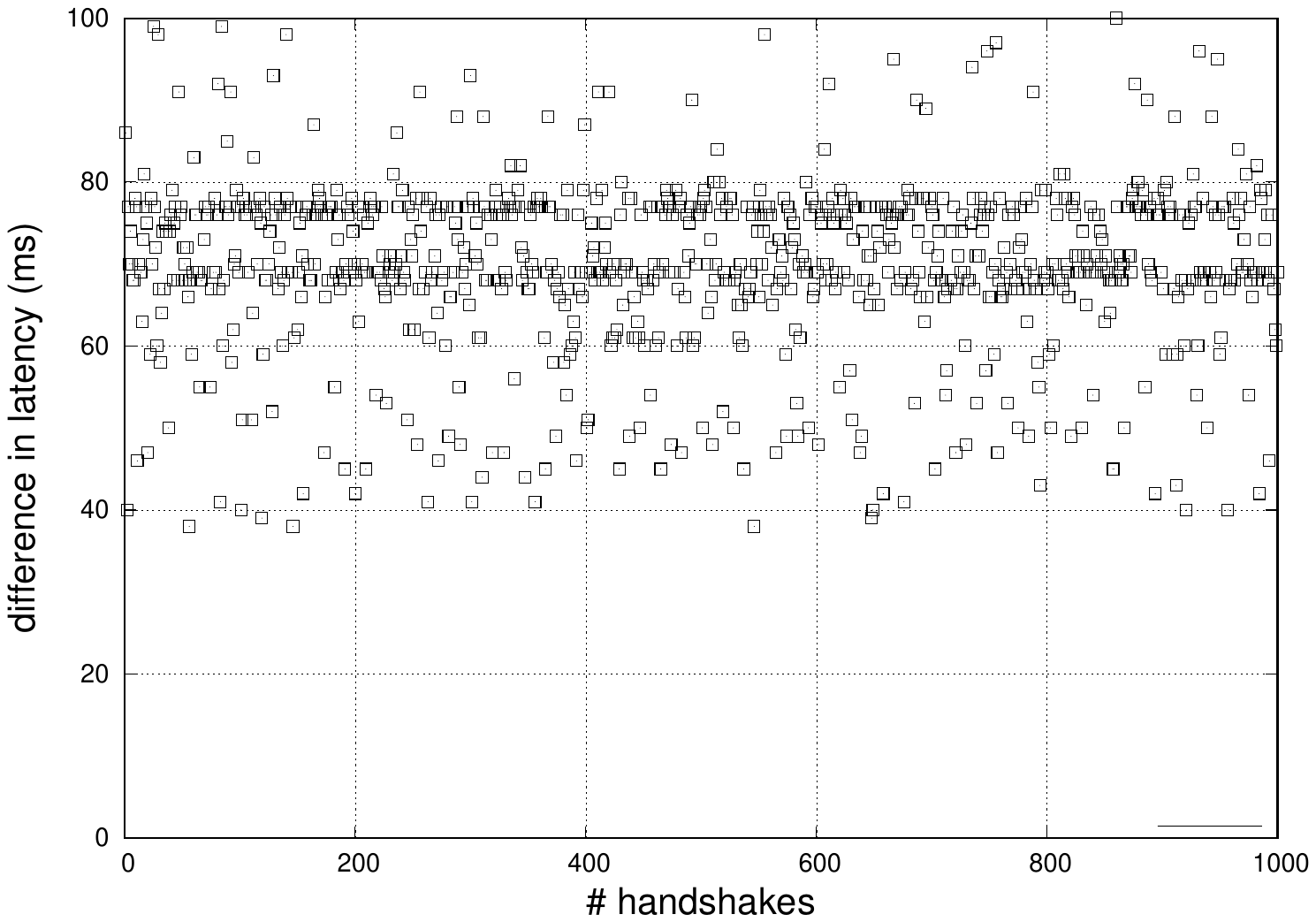}
        \caption{}
        \label{fig:two-way-measures}
    \end{subfigure}
    \begin{subfigure}{0.5\textwidth}
        \includegraphics[width=\columnwidth]{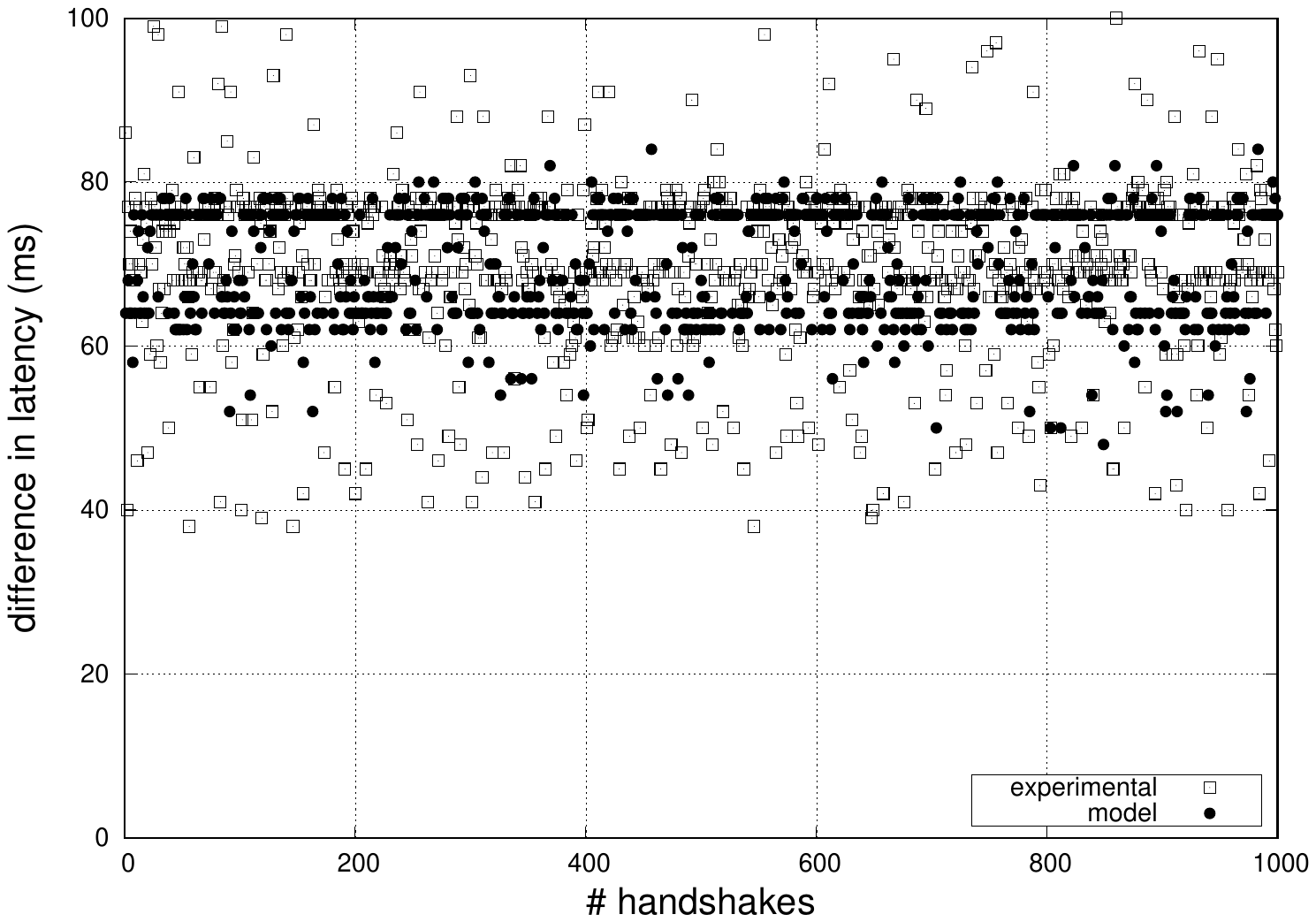}
        \caption{}
        \label{fig:two-way-model}
    \end{subfigure}
\caption{(a) Experimental difference between the mutually authenticated handshake latency with VC and X.509 authentication modes and ECDSA signature algorithm, (b) Overlay of experimental values and model estimates.}
\label{fig:two-way-measures-model}    
\end{figure*}

\subsubsection{unilaterally authenticated handshakes} the left part of the Table \ref{tab:symmetric-latency} shows the average latency of unilaterally authenticated handshakes for different configurations of signature algorithms and authentication modes. 
As expected from the analysis of the results in Table~\ref{tab:identity-verification}, the  handshake with X.509 certificates is the fastest under each signature algorithm. The handshakes with VCs and DIDs using RSA and ECDSA are very close, while with EdDSA the performance is again slightly better in favor of DID over VC mode. Although EdDSA is the slowest algorithm for signature verification algorithm, the RSA signature generation is expensive and this affects the overall latency of the handshakes.

Note that the analytical model discussed in Section~\ref{sec:model} well estimates the average latency with VC and DID authentication modes starting from the latency of the original handshake. Figure~\ref{fig:one-way-measures} depicts the difference between the latency experienced with VC and X.509 authentication modes. There is a concentration of points in the interval between 30 and 40ms, the experimental average is about 36 ms.
Given 1000 new experimental values for $T_V$, $T_C$ and $T_D$, we calculated $\Delta_V$ with equation~(\ref{eq:delta_v}). Figure \ref{fig:one-way-model} overlays the points in Figure~\ref{fig:one-way-measures} with the values of $\Delta_V$. The experimental and the model values overlap, meaning that equation~\ref{eq:hv-1} is a good estimate of the handshake latency with VC authentication mode. 
There is the same overlap with RSA and EdDSA signature algorithms and for the DID authentication mode, meaning equation~\ref{eq:hv-1} and \ref{eq:hd-1} are both good estimate of the handshake latency with VC and DID authentication modes for any signature algorithm.

\subsubsection{mutually authenticated handshakes} the right part of the Table \ref{tab:symmetric-latency} shows the average latency of handshakes for different configurations of signature algorithms and authentication modes. Client and server use the same authentication mode in these tests.
The same considerations done for the unilaterally authenticated handshakes applies to mutually authenticated handshakes. 
The use of X.509 authentication mode provides the lowest latency for all signature algorithms. The SSI handshakes suffer from an additional delay due to the need of authenticating the DLT node before the client and server can resolve the DID of the respective endpoint. 
As discussed in Section~\ref{sec:identity-verification}, the SSI handshake results are the worst case. With reference to the equation~\ref{eq:delta_v} and \ref{eq:delta_d},  the factor $T_D$ adds latency to the SSI handshakes and relevant amount of this delay, about 66 ms, is due to the secure channel setup with the DLT node. However, the solutions described in Section~\ref{sec:security} to reduce this delay can be applied to improve the performance of SSI handshake with both VC and DID authentication modes.
In any case, in the current setup, the best performance of the SSI handshake is with the ECDSA signature algorithm. 

Note again that the analytical model discussed in Section~\ref{sec:model} well estimates the average latency with VC and DID authentication modes starting from the latency of the original handshake. Figures \ref{fig:two-way-measures} and \ref{fig:two-way-model} show the validity of equation (\ref{eq:hv-2}) in the case of a mutually authenticated handshake with VC authentication mode. There is the same overlap with RSA and EdDSA signature algorithms and for the DID authentication mode, meaning equation~\ref{eq:hv-2} and \ref{eq:hd-2} are both good estimate of the mutually authenticated handshake latency with VC and DID authentication modes for any signature algorithm.

\subsubsection{hybrid handshakes} Table~\ref{tab:asymmetric-latency} shows the average latency of hybrid handshakes. These results maintain the same trends shown in the right part of Table~\ref{tab:symmetric-latency}. The hybrid handshake with ECDSA is the fastest, and it is the slowest with RSA.
Again, there is not much advantage in choosing VC over DID authentication mode, except in the case of EdDSA, where DID mode is 8\% faster than VC mode. Note that equations~\ref{eq:hybrid-v} and \ref{eq:hybrid-d} well estimate the latency in Table~\ref{tab:asymmetric-latency} using the values in Table~\ref{tab:identity-verification} and Table~\ref{tab:symmetric-latency} as inputs.

\section{Conclusion and future works}

With the aim of contributing to the adoption of SSI in large-scale IoT systems, this paper has presented a brand-new design of the TLS 1.3 handshake protocol using VC and DID authentication modes, while maintaining interoperability with the original TLS 1.3 handshake. Notably, the new protocol retains all the security features of the original one. The open source implementation of the novel TLS 1.3 handshake extensions and messages in OpenSSL allowed the authors to experimentally prove the feasibility of the proposed protocol. In addition, the experimental results have confirmed the validity of the analytical model developed to estimate the performance of SSI and hybrid handshakes with respect to the performance of the original one. 
The evaluation of the model and thus of the factors influencing the performance suggests the need to research a faster solution for the secure resolution of DIDs. The SSI and hybrid handshakes with both VC and DID authentication modes will benefit from this progress. A second important area of research is to find efficient ways to verify revoked identities during the handshake. A local revocation list approach can be adopted with all three means of authentication (i.e. certificate, VC and DID). 
In SSI handshakes with VC authentication mode an endpoint can access the DLT to retrieve the revocation status list stored by the Issuer~\cite{VC} or leverage an OCSP-like online service. 
Different considerations apply to the DID authentication mode. The DID resolution process returns either the DID Document with the identity public key of the other endpoint or a revocation proof if the owner has deactivated the self-issued DID. The revocation check is implicit in the DID resolution. This is a potential advantage of the SSI and hybrid handshakes that deserves further investigation. 

\newpage

\bibliographystyle{IEEEtran}
\bibliography{biblio}

\end{document}